\documentclass[notitlepage,floats,superscriptaddress,nofootinbib]{revtex4-1}

\usepackage{graphicx}
\usepackage[displaymath, mathlines]{lineno}

\usepackage[utf8]{inputenc} 
\usepackage[T1]{fontenc}    
\usepackage{hyperref}       
\usepackage{url}            
\usepackage{booktabs}       

\usepackage{nicefrac}       
\usepackage{microtype}      

\usepackage{mathtools}
\usepackage{mathrsfs}
\usepackage{commath}
\usepackage{bm}
\usepackage{amsfonts,amsmath,amssymb,amsthm}
\usepackage{color}
\usepackage[usenames,dvipsnames]{xcolor}
\usepackage{subfigure}
\usepackage{dsfont}
\usepackage{enumitem}
\usepackage{qcircuit}

\newcommand{\ket}[1]{| {#1} \rangle}

\newcommand{\ii}{\mathrm{i}}

\renewcommand{\a}[1]{\hat{a}_{\bm{#1}}}
\newcommand{\ad}[1]{\hat{a}_{\bm{#1}}^\dagger}

\newcommand{\phih}{\hat \phi}

\newcommand{\phin}{\hat \phi_{\bm n}}

\newcommand{\phinn}{\phi_{\bm n}}

\newcommand{\pin}{\hat \pi_{\bm n}}
\newcommand{\bn}{b_{n}}

\newcommand{\pih}{\hat \pi}
\renewcommand*\d[2][]{%
	\mathrm{d}%
	\ifx\relax#1\relax\else
	\rule{-0.02em}{1.5ex}^{#1}\rule{0.08em}{0ex}\!
	\fi
	#2\,
}
\newcommand{\kn}{\bm k_{\bm n}}
\newcommand{\on}{\omega_{\bm n}}

\begin{document}

\title{Learning to Utilize Correlated Auxiliary Noise: A Possible Quantum Advantage}


\author{Aida Ahmadzadegan}
\email{ahmadzadegan.aida@gmail.com}
\affiliation{Department of Applied Mathematics, University of Waterloo, Waterloo, Ontario, N2L 3G1, Canada}
\affiliation{Perimeter Institute for Theoretical Physics, Waterloo, Ontario N2L 2Y5, Canada}

\author{Petar Simidzija}
\affiliation{Department of Physics and Astronomy, University of British Columbia, Vancouver, British Columbia, V6T 1Z4, Canada}

\author{Ming Li}
\affiliation{Cheriton School of Computer Science, University of Waterloo, Waterloo, Ontario, N2L 3G1, Canada}

\author{Achim Kempf}
\affiliation{Department of Applied Mathematics, University of Waterloo, Waterloo, Ontario, N2L 3G1, Canada}
\affiliation{Perimeter Institute for Theoretical Physics, Waterloo, Ontario N2L 2Y5, Canada}
\affiliation{Institute for Quantum Computing, University of Waterloo, Waterloo, Ontario, N2L 3G1, Canada}
\affiliation{Department of Physics and Astronomy, University of Waterloo, Waterloo, Ontario, N2L 3G1, Canada}

\begin{abstract}
This paper has two messages. First, we demonstrate that neural networks that process noisy data can learn to exploit, when available, access to auxiliary noise that is correlated with the noise on the data. In effect, the network learns to use the correlated auxiliary noise as an approximate key to decipher its noisy input data. Second, we show that, 
for this task, the scaling behavior with increasing noise is such that future quantum machines could possess an advantage. In particular, decoherence generates correlated auxiliary noise in the environment. The new approach could, therefore, help enable future quantum machines by providing machine-learned quantum error correction.

\end{abstract}
\maketitle

\section{Introduction}
\label{intro}
Our first aim is to show that neural networks that learn a task on noisy data, such as, e.g., image classification, can simultaneously also learn to improve their performance by exploiting access to separate noise that is correlated with the noise in the data, when such auxiliary correlated noise is available. In effect, the network learns to improve its performance by implicitly using the auxiliary correlated noise to subtract some of the noise from the data. 

This new approach of `Utilizing Correlated Auxiliary Noise' (UCAN), has potential applications, for example, whenever noise arising in a measurement is correlated with noise that can be picked up in a vicinity of the measurement. 
The UCAN approach can also be applied in scenarios where the noise is added intentionally, for example, for cryptographic purposes. In the cryptographic case, the UCAN setup is essentially a generalized one-time-pad protocol \cite{onetimepad,Pirandola2019,Sergienko2018} in which the auxiliary noise plays the role of an approximate key that is correlated with the exact key that is represented by the noise on the data. In effect, the network uses the approximate key represented by the auxiliary noise to decipher the noisy data. 

The novel UCAN approach is, therefore, not primarily concerned with traditional denoising, see, e.g., \cite{NIPS2008_3506, PerezCisneros2016,Hinton_Imagenet, Bishop1995, classnoisycnn,CNN_noisy_image,Vincent_Pascal_2008,Vincent_Pascal_2010}, but is instead concerned with new opportunities for neural networks that arise in the event of the availability of correlated auxiliary noise. However, we will here not dwell on the range of possible conventional applications  from scientific data taking to signal processing and cryptography.  

Instead, our aim here is to provide a proof of principle of the UCAN approach on classical neural networks, in order to motivate to next apply the UCAN approach to quantum and quantum-classical hybrid neural networks. There, one application could be to the main bottleneck for quantum computing technology, the process of decoherence. This is because decoherence consists of the generating of correlated auxiliary noise in degrees of freedom in the immediate environment of the physical qubits. 
The challenge would be to try to access some of those quantum degrees of freedom and to machine-learn, in a UCAN manner, to re-integrate part of the leaked quantum information into the quantum circuit. This could yield a novel form of machine-learned quantum error correction that is not based on traditional quantum error-correction principles such as utilizing redundant coding or topologiocal stability but that instead tries to access environmental degrees of freedom to re-integrate previously leaked quantum information into the circuit. 

In the present work, we  aim to lay the ground work by demonstrating a proof of principle on classical machines. 
To this end, we here demonstrate the feasibility of the utilization of correlated auxiliary noise, i.e., of UCAN, through the intuitive example of convolutional neural networks that classify images. In particular, with a view to prospective applications to quantum noise, we here determine the scaling of the efficiency of the UCAN method as either the level of the noise, the dimensionality of the noise or the complexity of the noise are increased. We find that as the magnitude of the noise is increased, the efficiency of the UCAN approach increases. The efficiency becomes optimal in the regime where the magnitude of the noise is close to the threshold where the noise starts to overwhelm the network, i.e., where the performance of the network without UCAN would drop steeply. Further, we find that also as the dimensionality of the function space from which the noise is drawn is increased, the efficiency of the UCAN approach generally increases. Crucially, we also find that as the complexity of the noise is increased, the capacity of a neural network to use UCAN can easily be exhausted on classical computers. 

As we will discuss on theoretical grounds, this offers a potential advantage for quantum computers over classical computers in UCAN-type applications. The advantage could arise from the ability of quantum computers to store, and quickly draw from, extraordinarily complex probability distributions, even when operating only on a relatively small number of qubits. Regarding the possibility of using UCAN for machine-learned quantum error correction that we mentioned above, it is clear that only a quantum or quantum-classical hybrid neural network with quantum (i.e., non-measurement) access to environmental degrees of freedom can possibly achieve the UCAN task of re-integrating leaked quantum information from the environment into the quantum circuit.

For references on quantum computing, communication,  cryptography and error correction, see e.g.,  \cite{Gyongyosi2019, Pirandola2019, Sergienko2018,Devitt_2013}.

\section{Application of the UCAN approach to CNNs}

We begin with a concrete demonstration of UCAN on classical computers. While UCAN should be applicable to most neural network architectures, we here demonstrate the UCAN approach by applying it to image classification by convolutional neural networks (CNNs). 

To this end, we choose the standard Fashion-MNIST $28\times 28$ pixel grey level image data set and we add around the image, by zero-padding, a rectangular rim of black pixels which we refer to as a `bezel'.
We choose the bezel to be 6 pixels wide so that the number of pixels in the bezel around the image roughly matches the number of pixels in the image itself. We will refer to an image together with its bezel as a `panel', which has $40\times40$ pixels. First, we add noise only to the image part of the panels. The image classification performance of a CNN trained on these noisy panels correspondingly diminishes. We then examine to what extent the CNN can recover part of the noise-induced drop of its image classification performance when trained and tested with panels that possess noise on the image as well as noise on the bezel that is correlated with the noise on the image. 

Concretely, we generate three sets of labeled data. One set, A, consists of the original set of labeled MNIST images, with the black bezel added. The second set of labeled data, B, consists of the same set of labeled images with noise added only to the images. The third set of labeled data, C, consists of the same set of labeled images but with noise added to both the images and their bezels, with the image noise and the bezel noise generated so as to be correlated. 

We then train CNNs of identical architecture with the three sets of data and compare their image classification performance on the noisy images. We find that after the image classification performance drops from A to B, as expected, it increases again with C. This means that a CNN trained with noisy images with noisy bezel can outperform a CNN with the same architecture but trained on the noisy images with a noiseless bezel. This demonstrates that CNNs can be trained to use access to correlated noise on the bezel to improve their image classification performance by implicitly subtracting some of the noise from the image. 

The amount of performance recovery from B to C, as a fraction of the initial performance drop from A to B, may be called the efficiency of the UCAN method in the case at hand. In our experiments we explored how this efficiency depends on the level of the noise as well as on the dimensionality and the complexity of the noise. We will now discuss how we generate these varying types of noise. 

\subsection{Method to generate correlated noise of varying level, dimensionality and complexity}\label{noisegen}

{\bf Noise-to-Signal ratio.} \quad We increase the noise level, i.e., the noise-to-signal ratio, by increasing the noise amplitude range relative to the amplitude range of the pixels of the clear image. The brightness values of the clear image are ranging in the interval from zero (black) to one (white). We therefore lift and compress the brightness values of the clear image (and bezel) pixels to a suitable smaller range so that after the noise (whose amplitudes are allowed to take positive and negative values) is added, the brightness values of the noisy image and bezel is ranging again between zero and one.   

{\bf Dimension of the noise space.} \quad In addition to varying the noise-to-signal ratio, we are also varying the dimension of the space from which the noise is drawn. 
The dimension of the space of panels of size $40\times 40$ is $1600$. We choose a set of $N<1600$ basis vectors in that space and we then generate the noise as a linear combination of these noise basis vectors with coefficients drawn from a Gaussian probability distribution. In order to explore the scaling of the efficiency of the UCAN approach when increasing the dimensionality of the vector space from which the noise is drawn, we find that choosing the number, $N$, of noise basis functions to be either $5^2=25$ or $15^2=225$ or $22^2=484$ suffices to show the trend. (As shown in Sec.\ref{squares}, the squares arise when constructing the basis functions as the product of an equal number of Fourier modes in the $x$ and $y$ directions.) 

{\bf Noise complexity.} \quad  In order to vary the complexity of the noise, we choose the noise basis vectors such that the pixel pattern that they represent is either of low or high algorithmic complexity, i.e., such that it is either relatively easy or relatively hard to learn for a machine such as a neural network. On the notion of algorithmic complexity, see, e.g., \cite{Li2019}. In order to generate relatively low complexity noise, we choose as the basis vectors those pixel patterns that correspond to the first $5^2$, $15^2$ or $22^2$ sine functions of the discrete Fourier sine transform of the full image with bezel. Recall that sine functions are of low algorithmic complexity as they can be generated by a short program. In order to generate relatively high complexity noise, we span the noise vector space using $5^2$, $15^2$ or $22^2$ basis vectors that correspond to pixel patterns that approximate white noise. Recall that white noise is algorithmically complex. Correspondingly, it should become harder for a CNN to learn and utilize the more complex noise. 
Indeed, as the experimental results discussed in Sec.\ref{exp} show, the level of noise complexity that we can achieve by the above noise generating method is sufficient to reach the limit of noise complexity that the network architecture which we use in our experiments can accommodate for the purpose of UCAN. 

\begin{figure}[h]
    \centering
    \includegraphics[width=0.3\linewidth]{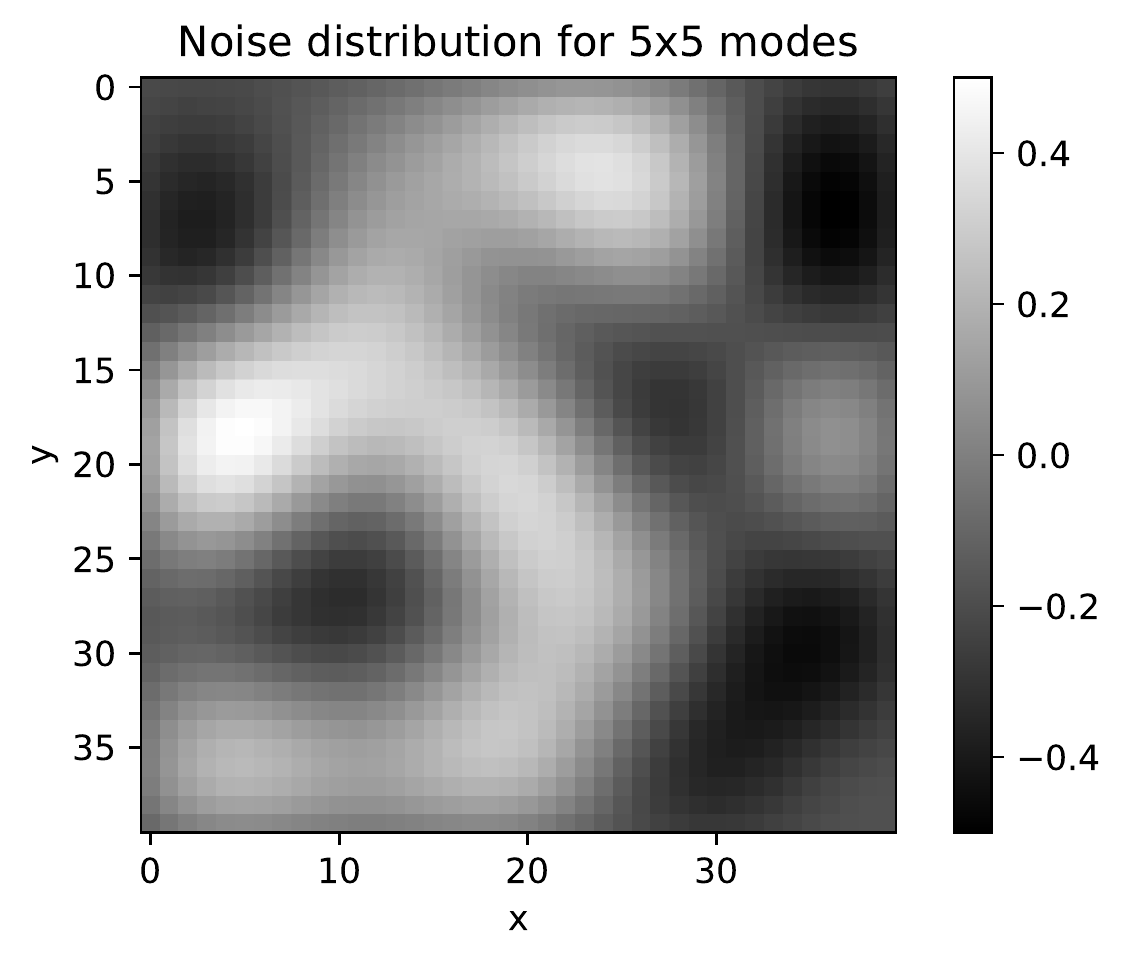}
    \includegraphics[width=0.3\linewidth]{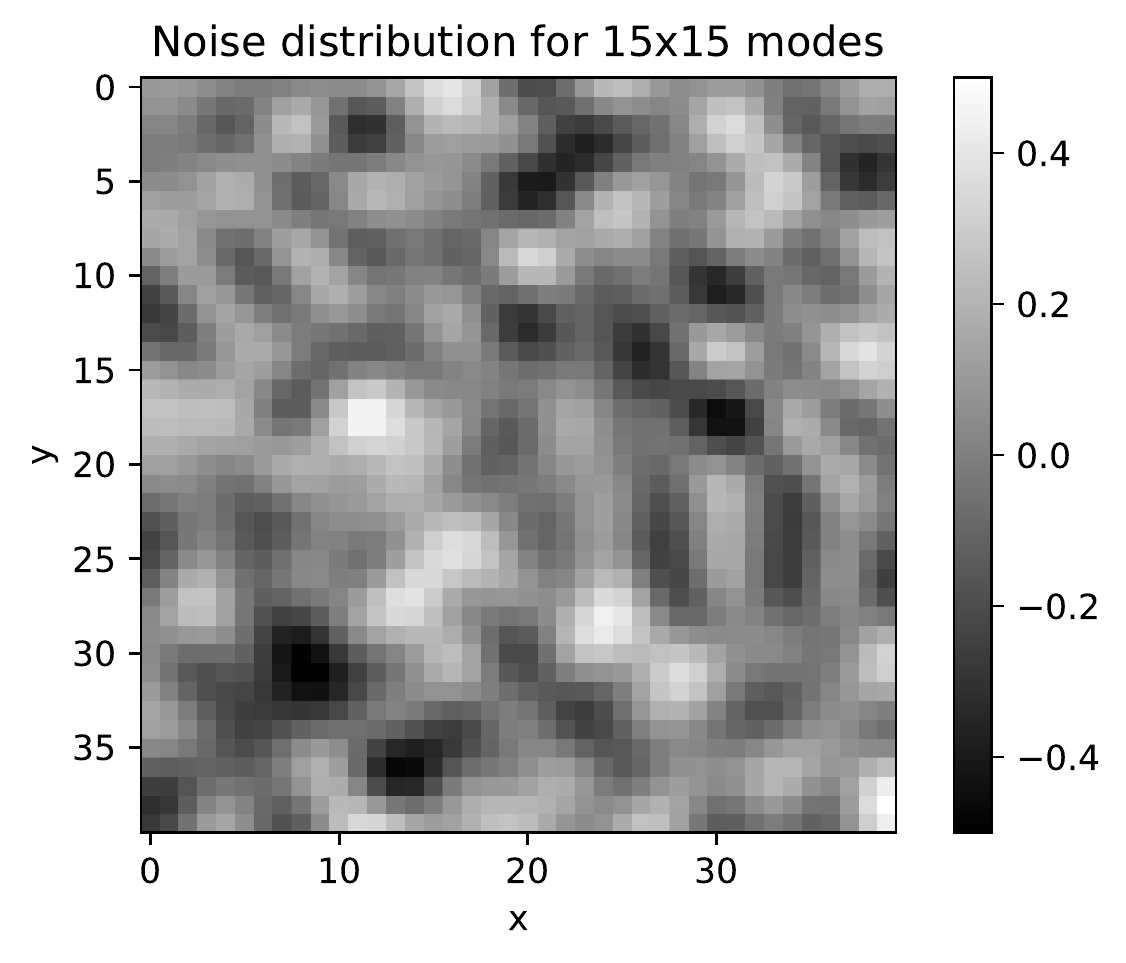}
    \includegraphics[width=0.3\linewidth]{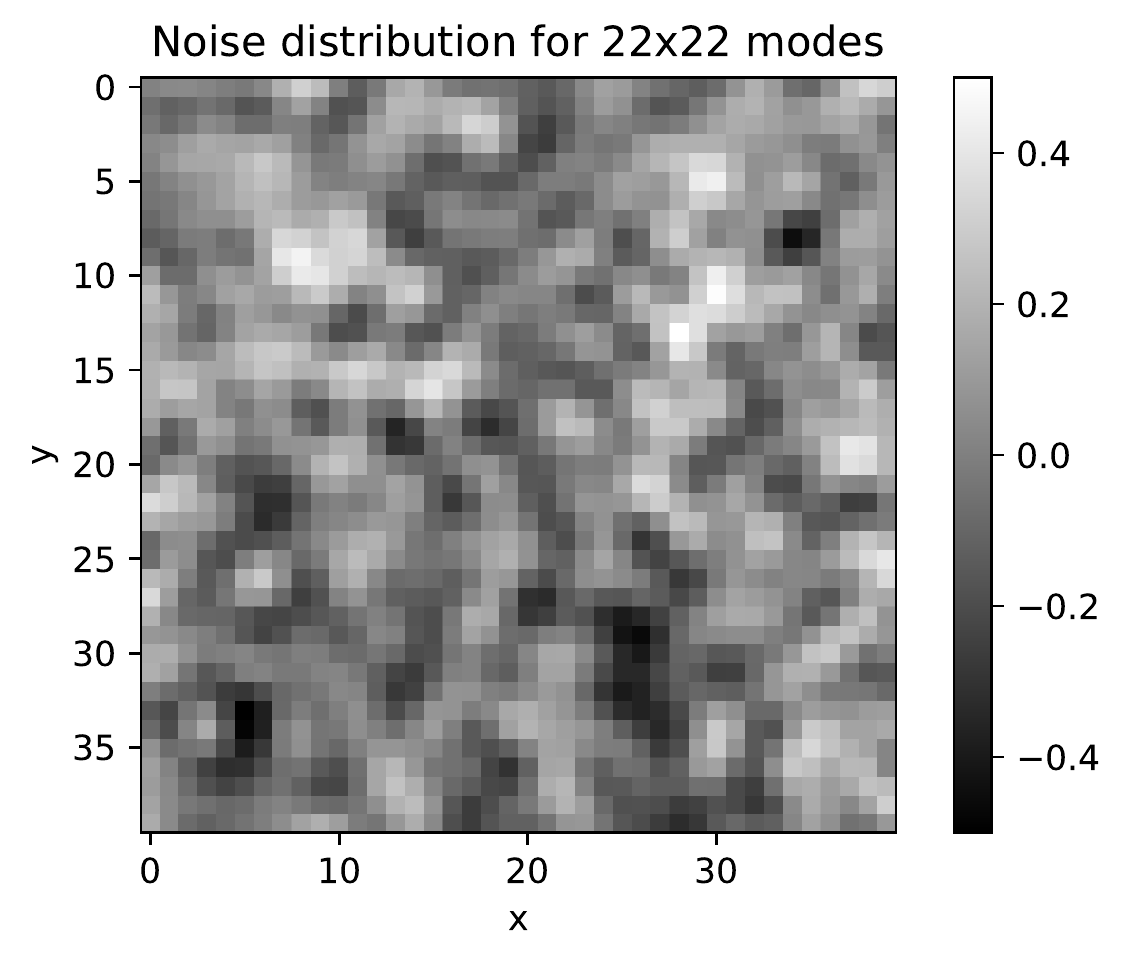}
    \caption{Examples of low complexity noise panels drawn from noise spaces of dimensions $(5\times5)$, $(15\times15)$, and $(22\times22)$ respectively.}
    \label{noise2}
\end{figure}

\begin{figure}[h]
    \centering
    \includegraphics[width=0.3\linewidth]{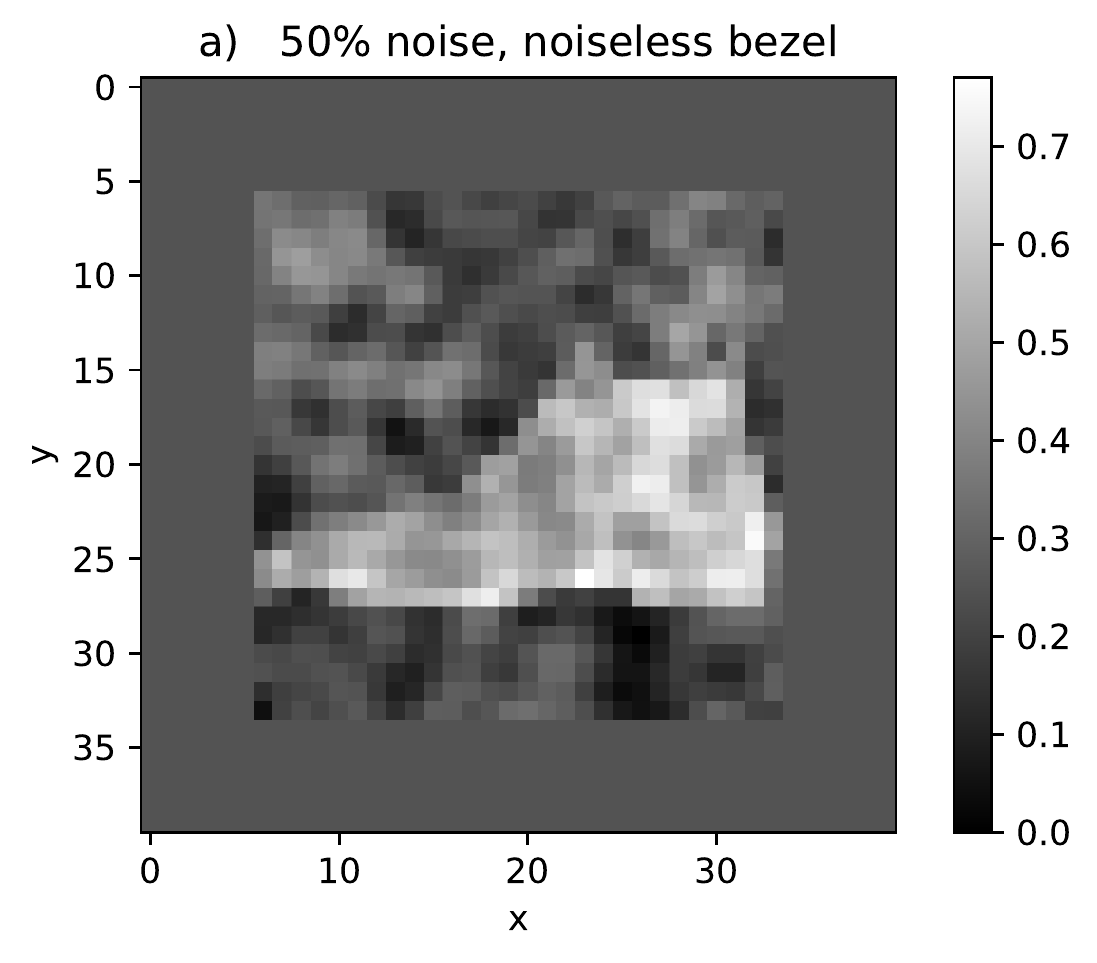}
    \includegraphics[width=0.3\linewidth]{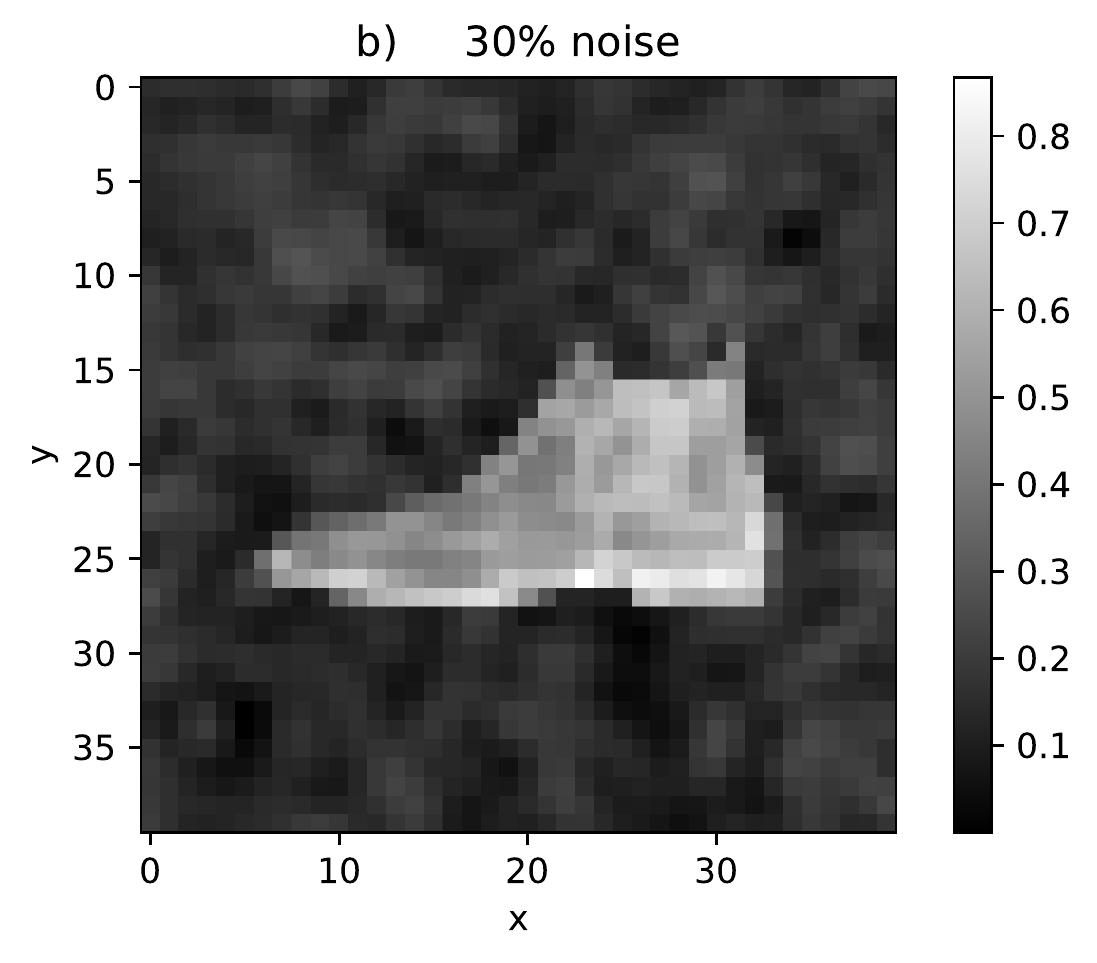}
    \includegraphics[width=0.3\linewidth]{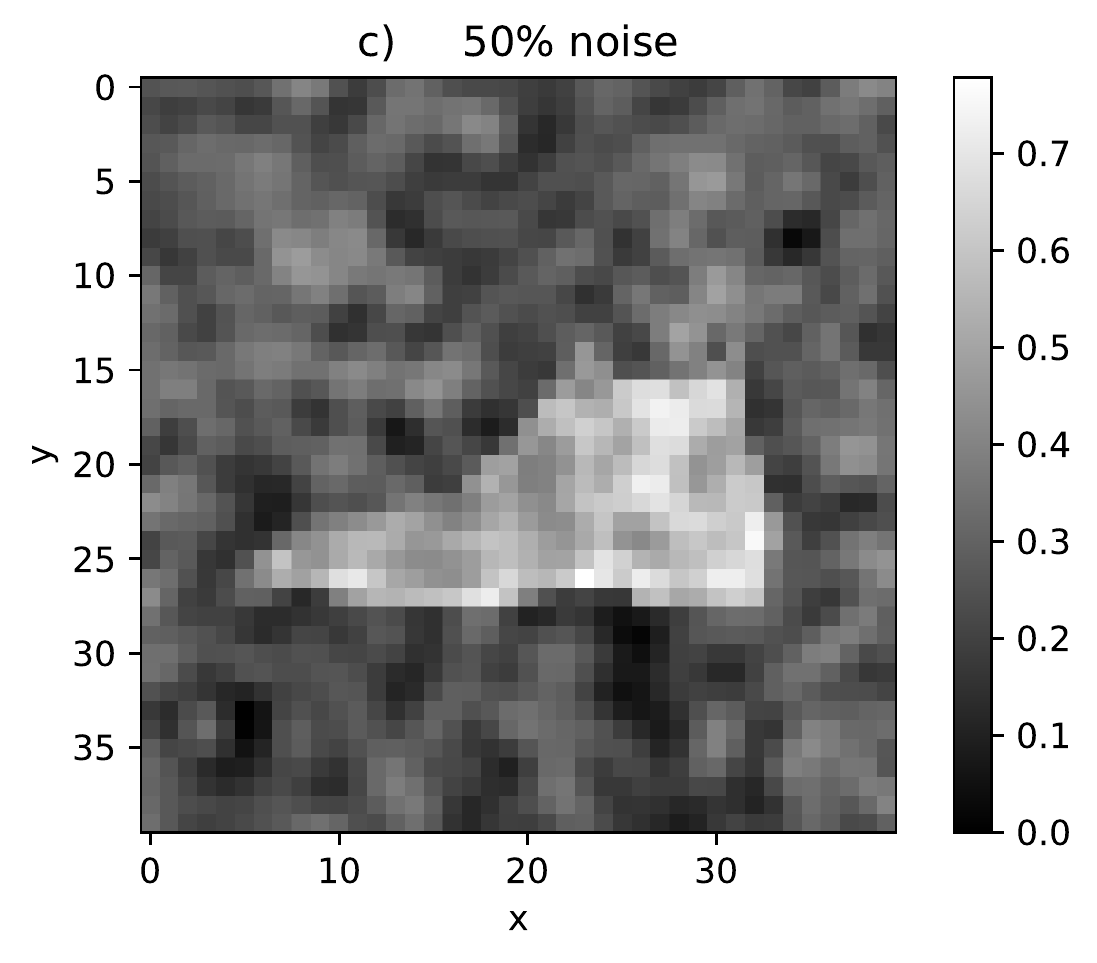}
    \includegraphics[width=0.3\linewidth]{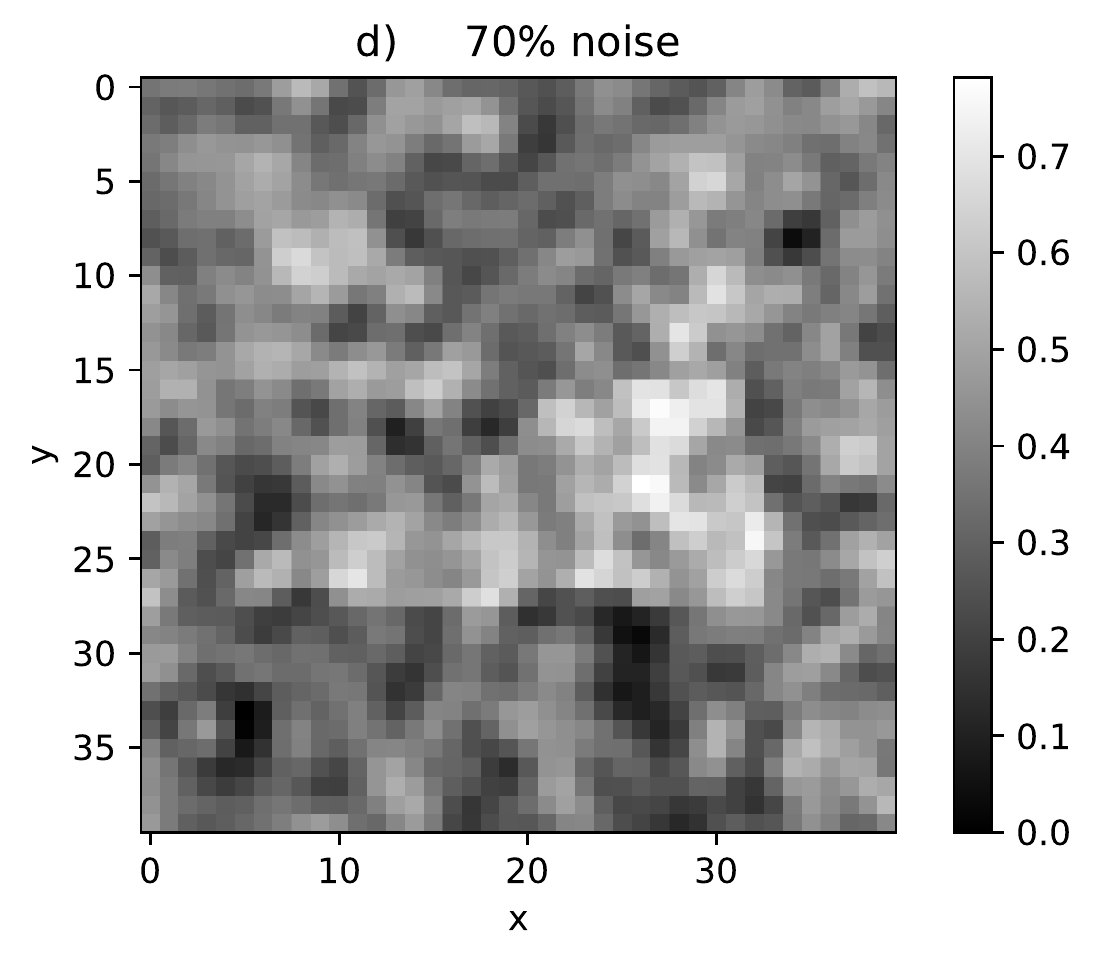}
    \includegraphics[width=0.3\linewidth]{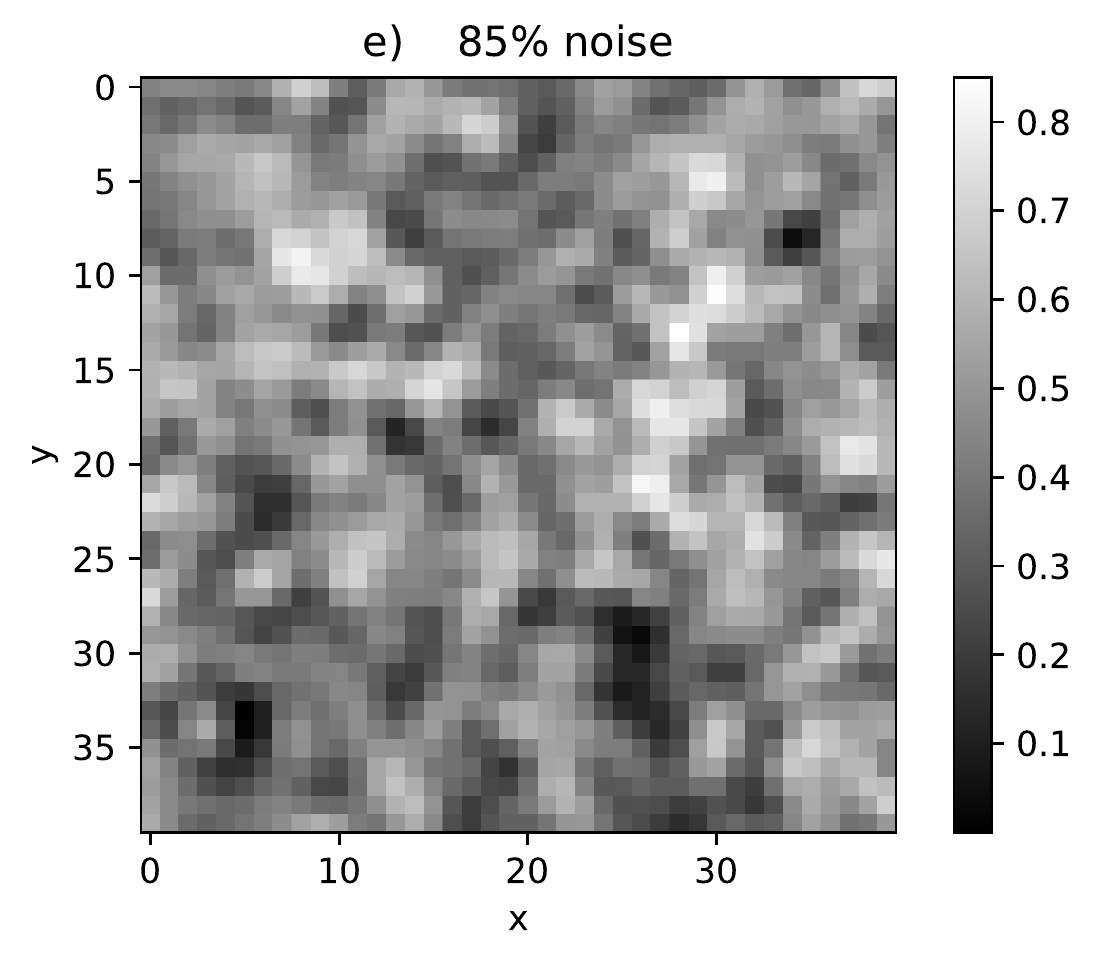}
    \includegraphics[width=0.3\linewidth]{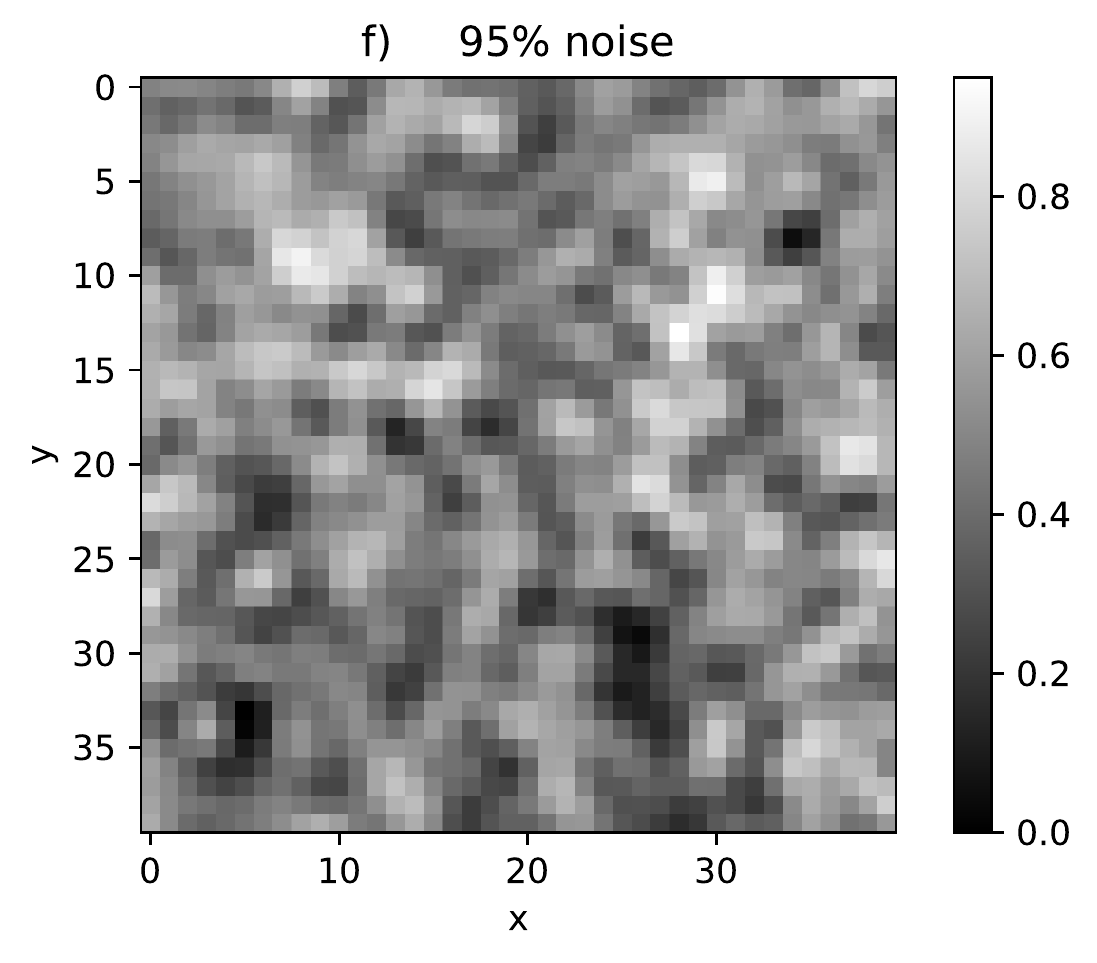}
    \caption{a) Processed image from Fashion-MNIST dataset with added 6 pixel-width initially black bezel. All pixels are rescaled to an interval of $[0.25, 0.75]$ (therefore the grey bezel has pixel values of $0.25$). Finally, noise of amplitude interval $[-0.5, 0.5]$ is added to the $28\times28$ image.  Panels b) to f) show examples where 30\%, 50\%, 70\%, 85\%, and 95\% of the panel is noise, respectively.}
    \label{noiseperc}
\end{figure}

Fig.\ref{noise2} shows examples of panels of relatively low complexity noise drawn from noise spaces of increasing dimension.  
Fig.\ref{noiseperc} shows panels of a noisy image and bezel with increasing noise-to-signal ratios. The noise-to-signal ratio is increased until the image is no longer classifiable by human perception. In the experiments, we increase the noise-to-signal ratio until the networks classify no better than chance.  

{\bf Quantum perspective.} \quad  In Sec.\ref{quantumUCAN} and in the Appendix, we will show that the above method for generating noise can also be viewed as a classical simulation of the quantum noise generated by a quantum system such as a quantum field in a suitable quantum state. For example, the generating of low-complexity noise as a linear combination of Fourier sine functions with Gaussian distributed coefficients can be viewed as accurately simulating the vacuum fluctuations of a bandlimited Klein-Gordon (KG) quantum field in two dimensions. On quantum field fluctuations, see, e.g., \cite{Liddle2000, Mukhanov2005, MukhanovWinitzki2007,Birrell1984,Achim_Rob_minUR,JasonAchim_bandlimited,Achim_unsharpcoor_PRL,AidanAchimRobert2017,AidaAchimRobert2013}. In Sec.\ref{quantumUCAN}, we will discuss the scaling of the noise complexity that could be achieved in the UCAN context through the use of generic highly entangled states on a quantum machine. 

\section{Experiments}
\label{exp}
\subsection{Experimental setup}
In this section, we detail our implementation of the new UCAN scheme in convolutional neural networks. We use the slightly modified version of CNN architecture given in \cite{CNN4} which contains three convolutional layers and two fully connected layers. Full details regarding our network architecture, training, and evaluation are provided in the Appendix.

In brief, our training data sets are generated from the set of labeled $28\times 28$ pixel Fashion-MNIST images \cite{fashion}. The data set contains 10 different types of fashion items, i.e., if a CNN performs at the level of 10\% accuracy then it classifies no better than chance. The data set consists of 10k test images and 60k training images which we divided into a 50k training and a 10k validation set. The Fashion-MNIST images are in grey scale with the pixel values originally ranging between 0 and 255, including both bounding values. We re-scale these values to the interval $[0,1]$.

To obtain our data sets of type A, we add to the images a black bezel of 6 pixel width by zero-padding. We obtain data sets of type B by adding noise only to the image, as, e.g., in Fig.\ref{noiseperc}a, and data sets of type C by adding noise to both the image and the bezel, as, e.g., in Figs.\ref{noiseperc}b-f.

\subsection{Generating low and high complexity noise} \label{squares}
In order to generate noise with relatively low algorithmic complexity, we construct a basis of the noise space by using the orthogonal sine functions of the Fourier sine transform of functions defined on the square $[0,L]\times[0,L]$: 
\begin{linenomath}
\begin{equation}
\bn(\bm x)=\frac{2}{L}
\sin\left(\frac{n_x\pi x}{L}\right)\sin\left(\frac{n_y\pi y}{L}\right)
\end{equation}
\end{linenomath}
Here, $n=(n_x,n_y)$ is a pair of positive integers that label the choice of basis function. Each basis function $b_n(x)$ yields a $40\times 40$ panel, $P_n$, by evaluating the basis function on the grid of integers:  $P_n:=[b_n(m_1,m_2)]_{m_1,m_2=1}^{40}$. Each such panel serves as a basis vector in the space of panels from which we draw the noise. In order to avoid needlessly small amplitudes near the boundary (due to the vanishing of all sines there), we choose $L$ slightly larger than $40$, at $44$. We then generate each noise panel, say $r$, as a random linear combination of the $N=M^2$ basis panels obtained from the $M$ first sine functions in the $x$ and $y$ directions. The pixel values of $r$ are
\begin{linenomath}
\begin{equation}\label{eq:Fourier_expansion}
    r_{(m_1,m_2)}:=\sum_{n_1,n_2=1}^M  g_{(n_1,n_2)} P_{(n_1,n_2)}(m_1,m_2) ~~~\mbox{where}~~~m_1,m_2 = 1,...,40,
\end{equation}
\end{linenomath}
where we choose the coefficients $g_{(n_1,n_2)}$ from  Gaussian probability distributions\footnote{We choose the width of the Gaussians to be $\omega(n_1,n_2):=1/\sqrt{(n_1^2+n_2^2)}$. This choice lessens the probability of large amplitudes of the coefficients of sines of short wavelength, leading to a pink noise spectrum. We choose these Gaussian distributions since, as discussed in detail in the Appendix, this choice also happens to exactly match the statistics of the quantum vacuum fluctuations of a neutral scalar Klein-Gordon quantum field.}. Examples of noise panels drawn from noise spaces of different dimensions, $N=M^2$, are shown in Fig.\ref{noise2}. 
Since we have 60k training and 10k test images, we need 70k such noise panels to add to our total 70k Fashion-MNIST dataset. In order to study the effect of increasing the dimension of the noise space on the performance of the network, we create three such data sets of 70k panels each, with the noise space of dimensions $N=5^2$, $15^2$, and $22^2$, respectively. 

In order to generate noise with high algorithmic complexity, we proceed exactly as above, except that we use as the basis of the space of noise panels not sine functions but instead panels of fixed approximate white noise. Each of the basis noise panels is generated by drawing for each pixel its grey level from a normal distribution. For later reference, let us note here that the so-obtained basis noise panels are generally not orthogonal, unlike the sine based base noise panels. The 70k noise panels are then generated each as a linear combination of these basis noise panels, with coefficients drawn from a Gaussian probability distribution and truncated so that the grey levels of the noise panel is in the range of $[-0.5,0.5]$. Analogous to the case of relatively low complexity noise, we generate also the sets of relatively high complexity noise panels by  linearly combining, with Gaussian-distributed random coefficients, either $N=5^2$, $15^2$, or $22^2$ basis noise panels. 

\subsection{Experimental results}

\begin{figure}[h]
    \centering
    \includegraphics[width=0.497\linewidth]{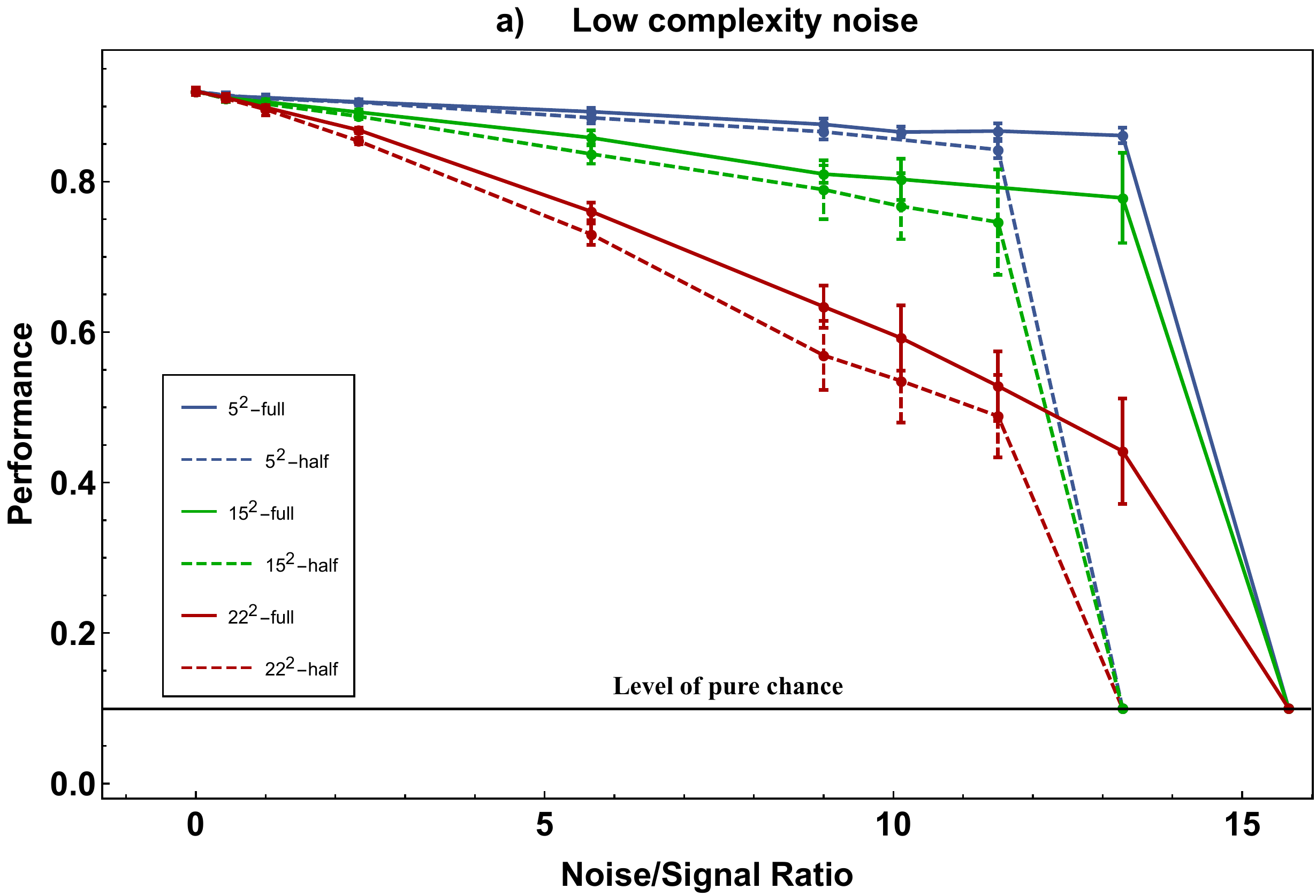}
    \includegraphics[width=0.497\linewidth]{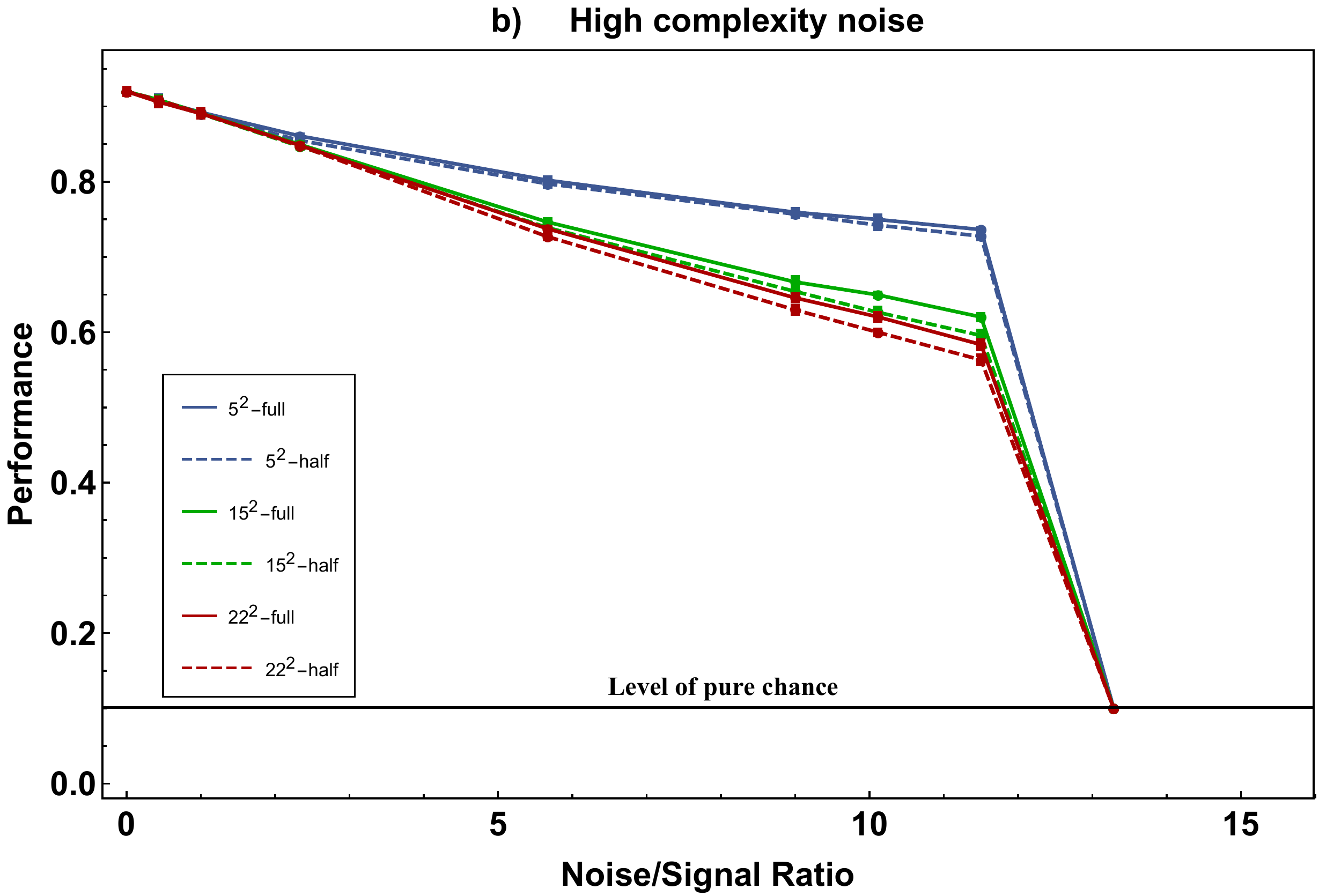}
    \caption{The test accuracy, i.e., the performance of the network on the test dataset, as a function of the noise-to-signal ratio. Notice that, since there are 10 different fashion items, a success rate of 10\% indicates that the network classifies at a rate that is equal to pure chance.}
     \label{testaccuracy}
\end{figure}

The experimental results, i.e., the performances of our convolutional neural networks as a function of the level, dimensionality and complexity of the noise are shown in Fig.\ref{testaccuracy}. The $y$-axis indicates the performance of the CNN and the $x$-axis denotes increasing levels of noise. The left panel, Fig.\ref{testaccuracy}a, shows the performance for noise of relatively low computational complexity, i.e., noise arising as linear combinations of basis noise panels that represent sine functions. The right panel, Fig.\ref{testaccuracy}b, shows the performance for noise of high computational complexity, i.e., for the noise that arises as linear combinations of basis noise panels that each represent approximate white noise. The blue, green, and red curves in Fig.\ref{testaccuracy}a,b represent the choice of $N=5^2,15^2$ or $22^2$ dimensions for the space of noise panels. 

The dashed lines represent the performance of the CNN on the data sets with the noise only on the image while the solid lines represent the performance of the CNN with the noise both on the image and on the bezel. Each data point has been calculated multiple times and the mean value together with its standard deviation in the form of error bars is plotted. The error bars on Fig.\ref{testaccuracy}b are there but they are small, as we will discuss below.

We begin our analysis of the experimental data with the observation that the curves show that, as the level of noise increases, the performance generally drops. 
In addition, we notice that on the noise-to-signal ratio axis, there are well-defined `cliffs' where the performance sharply drops to the level of 10\% and the network is no longer able to learn to classify better than chance. We also see that the performance drops as the dimensionality of the noise space is increased, i.e., from blue to green to red. As the complexity of the noise is increased, namely from Fig.\ref{testaccuracy}a to \ref{testaccuracy}b, the performance also drops - except for the red curves, i.e., except if the dimension of the noise space is highest. We will discuss this exception further below.

The most crucial observation, however, is that all the solid lines are above the dashed lines. This means that the CNNs were able to improve their performance due to UCAN, i.e., when they are given access to correlated noise on the bezel. In particular, we see in Fig.\ref{testaccuracy}a that the efficiency of UCAN, i.e., the gap between the dashed and solid lines of equal color, increases with increasing noise level. 
Most importantly, we observe that the cliff at which the performance of the network drops sharply is at a higher noise level for the solid lines, with UCAN, than it is for the dashed lines, without UCAN, i.e., without noise on the bezel. Concretely, we observe that there exists a special regime of noise-to-signal levels, here in Fig.\ref{testaccuracy}a around $14$. At that level of noise, a CNN without UCAN (dashed lines) cannot learn at all, i.e., its performance drops to 10\%, which is the performance level of pure chance. At the same level of noise, however, a CNN of the same architecture but with access to correlated auxiliary noise on the bezel (solid lines) learns to perform considerably well, here with performance levels from about 40\% to about 90\%, depending on the dimension of the noise space. The upshot is that UCAN possesses its highest efficiency in the regime of such high noise-to-signal ratio, where the network without UCAN starts to fail to learn at all. 

It is intuitive that the efficiency of UCAN is best in regimes of high levels of noise. This is because UCAN in effect reduces the network's rate of those misclassifications that are due to noise while, at low noise levels, most misclassifications of a CNN are not primarily due to noise. However, we can only expect the efficiency of UCAN to increase with increasing noise as long as the capacity of the network suffices to learn to utilize the correlations in the noise. Indeed, our experiments showed that the networks struggled to achieve UCAN efficiency in the regime of high noise complexity: in Fig.\ref{testaccuracy}b, the solid lines are barely above the dashed lines. This demonstrates that the UCAN approach can quickly exhaust a classical network's capacity. In Sec.\ref{quantumUCAN}, we will come back to this point in our discussion of the prospect of UCAN on quantum machines, which should possess a much higher capacity to represent complex correlations. 

Let us now also discuss why the error bars on Fig.\ref{testaccuracy}b are smaller than those on Fig.\ref{testaccuracy}a. Superficially, the reason is that the performance of the CNNs was more uniform in the case of the high complexity noise on the right. We conjecture the reason to be that the network, when trained on the low complexity noise data, succeeded to learn, to a varying extent, the algorithmically relatively simple long distance correlations between bezel and image noise that are due to the algorithmically relatively simple nature of the sine functions. In contrast, the CNNs appear to have consistently struggled to learn any correlations between the bezel and image noise in the case of relatively high noise complexity. 

Finally, let us discuss why the red curves are higher in Fig.\ref{testaccuracy}b than in Fig.\ref{testaccuracy}a. We expect the reason to be that the $22^2$-dimensional noise space on the left is spanned by sine functions that are orthogonal while $22^2$-dimensional noise space on the right is spanned by $22^2$ random white noise panels that are at random angles to another. This means that the noise space is more uniformly sampled for the red curves on the left than for those on the right, which leads to more predictability of the noise on the right and therefore to an advantage for the CNNs on the right. This phenomenon arises only for high-dimensional noise spaces where the directions of random basis vectors start crowding together.

\subsection{Correlation analysis}

So far, we discussed the efficiency of the UCAN method as a function of the the noise-to-signal-ratio, the noise dimensionality and the algorithmic complexity of the noise. We are now ready to discuss the performance of the UCAN method in terms of the correlations between the noise on the bezel and the noise on the image. 

We begin by noting that, since uncorrelated noise is of no use for UCAN, we chose all of the noise in our experiments to be perfectly correlated between the bezel and the image. If the noise on the bezel was known then, in principle, the noise on the image could be perfectly inferred. To see this, let us consider the simple case where the noise space is one dimensional, i.e., where all noise panels are a multiple of just one basis noise panel. In this case, knowing one pixel value anywhere, for example on the bezel, would imply knowing the noise everywhere. More generally, if the noise space is chosen to be $N$-dimensional, then knowledge of the grey level values of any $N$ pixels, e.g., $N$ bezel pixels, if the bezel has enough pixels, allows one to infer the noise everywhere, namely by solving a linear system of equations. Since the largest dimension of the noise space that we considered is $N=22^2=484$, while the bezel possesses a larger number of pixels, namely $B=40^2-28^2=816$, it is always possible to determine the noise on the image from the noise on the bezel, in principle. However, for a network to infer the image noise from the bezel noise, it would first need to determine the exact noise space. One challenge for the network is that while it is trained with a clear view of the noise on the bezel, its view of the noise on the images is obscured by the presence of the images. 

More importantly, some noise spaces are easier for a network to learn than others. 
For example, if the noise space is one dimensional, then the network needs to learn only one noise basis panel. If this panel is simple, e.g., if the grey level values follow a sine wave, then the panel is easier to learn than when the noise basis panel is of high algorithmic complexity, such as a panel of white noise. The challenge to the network increases as the dimensionality of the noise space is increased. Experimentally, as is clear when comparing Fig.\ref{testaccuracy}a and Fig.\ref{testaccuracy}b, it is indeed easy to overwhelm the network's limited capacity to benefit from UCAN by using basis noise panels of high algorithmic complexity. 

Our experiments have been limited, so far, to UCAN applied to CNNs for image classification. 
It should be very interesting to apply UCAN to other network architectures whenever auxiliary correlated noise is naturally available or can usefully be added, e.g., as the case may be, with RNNs for signal processing, or autoencoders for denoising. 

Independently of which suitable neural network architecture the UCAN method is applied to, we are led to conjecture that the efficiency of the UCAN method tends to increase as the amount of noise increases, and that the efficiency of UCAN is highest when the noise reaches the level at which the network without UCAN would start to fail to learn. We are also led to conjecture that when UCAN is applied to any neural network architecture, then even perfect correlations between the noise on the input signal and the auxiliary noise can easily be made sufficiently complex to exhaust the capacity of the network to learn to utilize these correlations. 

To support this conjecture, let us discuss to what extent the complexity of the noise could be increased. For example, in the case of the CNNs that we studied here, the noise panels do not need to be generated in the way we did, i.e., by linearly combining noise basis panels with independently distributed coefficients. Instead, in principle, the noise panels could be drawn from any probability distribution over the manifold $[0,1]^{40\times 40}$, i.e., over the $1600$-dimensional unit cube. Even if the pixel values are restricted to be $0$ or $1$, a generic and therefore highly complex probability distribution would require the specification of $2^{1600}\approx 10^{481}$ coefficients. This confirms,  in this example, that even if the noise that occurs in practical applications of UCAN is manageable for a suitable network, the complexity of the noise could easily be increased to exceed any network's ability to learn or store or draw from its probability distribution, at least if running on a classical machine. 

\section{Highly complex noise correlations and quantum UCAN}\label{quantumUCAN}

Quantum machines may offer advantages for neural networks using UCAN, especially in the regime of highly complex noise, as a quantum machine with $R$ qubits can store probability distributions described by a $2^R$-dimensional Hilbert space. 
Also, while classically it is generally prohibitively expensive to draw from high-dimensional probability distributions, quantum machines allow one to easily draw from any of its states' probability distribution, through measurement. Further, the quantumness provides a source of true randomness rather than approximate randomness, as the violation of Bell inequalities shows, see, e.g., \cite{BellTest1998,BellTest2017}. In addition, the quantum mechanical Hilbert space of probability distributions is richer than that of classical distributions, due to the additional dependence on the choice of measured observables. This in turn allows entangled states that violate Bell-type inequalities to describe correlations, say between noise on data and auxiliary noise, that could not arise from localized classical dynamics. 

Illustrating the generality of the quantum perspective is the fact that the noise that we used in the experiments here could also be generated by a quantum system, namely by a $2$-dimensional neutral massless Klein-Gordon quantum field (which is similar to one polarization of the quantized electromagnetic field) discretized to a $40\times 40$ grid. As we describe in detail in the Appendix, the low complexity noise that is generated using sinusoidal noise panels statistically exactly matches the quantum fluctuations of the amplitudes of the Klein-Gordon field in the vacuum state with a hard UV cutoff, i.e., with a bandlimit, \cite{kempf2004covariant,Achim_Rob_minUR,JasonAchim_bandlimited,Achim_unsharpcoor_PRL,AidanAchimRobert2017,AidaAchimRobert2013} determined by the dimension of the noise space. The entanglement entropy in the vacuum state obeys an area law and is correspondingly low, consistent with the fact that the noise here is of low algorithmic complexity. For the relationship between Shannon entropy (here in the form of von Neumann entropy) and algorithmic complexity, see, e.g., \cite{Li2019}. 
The statistics of the high complexity noise generated using white noise panels also matches the field's fluctuations, namely if the field's state is a suitable superposition of field-amplitude eigenstates. 

While a quantum field can, therefore, generate the noise that we considered in our experiments, it is also capable to generate noise of extremely higher complexity. In fact, it is known \cite{Page1993}, that any generic pure state is close to being maximally entangled and possesses close to maximum entanglement entropy between two equal size partitions of the system, such as here the bezel and the image. The almost maximal von Neumann entropies of the noise on the data and the auxiliary noise then imply a correspondingly almost maximal algorithmic complexity of the noise, illustrating the ability of quantum systems to efficiently store and draw from truly highly complex probability distributions.

\section{Outlook}

In classical contexts, whenever correlated auxiliary noise is available along with noisy data, or whenever correlated auxiliary noise can be usefully added, the UCAN approach for classical neural networks should be relatively straightforward to implement along the lines presented above.  

Let us, therefore, now focus on the question of the potential availability of correlated auxiliary quantum noise for applications of UCAN, with a view especially on quantum technologies.  
In the literature, there are indeed a few examples of uses of auxiliary quantum noise, although so far we know of none that is bringing the power of machine learning to bear, as is our proposal here. 

For example, in the quantum energy teleportation (QET) protocol \cite{HOTTA20085671,Koji-superadditive}, an agent invests energy into a local measurement of quantum noise and communicates the outcome to a distant agent who, on the basis of entanglement in the underlying medium or vacuum, uses this information to correspondingly interact with the agent's local quantum noise, enabling that agent to locally extract energy. Quantum energy teleportation has been generalized to aid in algorithmic cooling in quantum processors, \cite{Boykin3388,Rodr_guez_Briones_2017,Nayeli2017Alcool}. Also, for example, in quantum optics, see, e.g., \cite{Walls2007, Bachor2004}, the technique of ghost imaging is based on utilizing what is effectively correlated auxiliary classical or quantum noise, see, e.g., \cite{opticalimaging_pittman,Bornman2019}. 

Further, it was shown in \cite{Koji-superadditive} that in communication through a quantum field, access to correlated auxiliary quantum noise is always available to the receiver, due to the ubiquitous entanglement in quantum fields and that, in principle, this auxiliary noise is usable to increase the channel capacity. This suggests that classical or quantum implementations of UCAN on quantum machines could be useful, for example, to improve the classical or quantum channel capacity within or between quantum processors or quantum memory. In this case, the quantum noise on the data and the correlated auxiliary quantum noise would arise from the quantum fluctuations of the quantum field that is used for the communication, such as the electromagnetic field, or a quantum field of collective excitations such as the effective phononic field of ion traps \cite{trappedion}. In the context of superconducting qubits, see in particular also, e.g., \cite{HarmutNeven2019}.

Finally, there is the possibility that UCAN could be used as a method of machine-learned quantum error correction, as we mentioned in the introduction. 
Indeed, the decoherence of a quantum processor through interaction with its environment consists of creating correlated auxiliary quantum noise in the environment. The experimental challenge then, is to give a quantum neural network, or a quantum-classical hybrid neural network, access to some of the relevant `environmental' quantum degrees of freedom (which can also be located in the processor itself), as well as access to the quantum processor's noisy quantum output. The computational challenge would be to train the quantum network to undo some of the deleterious effects of the decoherence. 

While in our study of UCAN on classical machines above, the decoherence-induced quantum noise in the environment corresponds, of course, to the bezel noise while the noisy output of the quantum processor corresponds to the noisy image, the quantum network's architecture can be very different from that of a CNN. Nevertheless, if we can use our results for classical CNNs of above as guidance, we may speculate that a UCAN approach to machine-learned quantum error correction  (as compared to the traditional, scripted approach to quantum error correction that works well at low noise levels) may also work well in the regime of relatively strong noise or strong decoherence. Nevertheless, the quantum neural network will of course itself have to possess suitably low noise levels and it should be very interesting to determine corresponding threshold theorems, see, e.g., \cite{knill1998resilient}. 

For recent work on quantum machine learning and quantum neural network architectures, see, e.g.,  \cite{Benedetti_2019,Dunjko_2018,Biamonte2017,Ciliberto2018QuantumML,Schuld2014TheQF,Yang2020EntanglementbasedQD,QTensorFlow,Guill_JasonQML,QGNN,LtoL_QNN}. It should be very interesting to determine which  quantum neural network architectures are best suited for quantum UCAN applications, such as quantum machine-learned error correction.


$$$$
\bf Acknowledgements \rm
\bigskip\newline\noindent
This research was enabled in part by support provided by Compute Canada (www.computecanada.ca). AK is acknowledging support through the Discovery Program of the National Science and Engineering Research Council of Canada (NSERC), support through the Discovery Project Program of the Australian Research Council (ARC) and through two Google Faculty Research Awards. PS acknowledges  support through an NSERC Canada Graduate Scholarship. ML acknowledges support through the  Canada Research Chair and Discovery programs of NSERC.

\bibliographystyle{unsrt}
\bibliography{ref}

\newpage

\appendix

\section{Network architecture, training and evaluation details}

{\bf Network architecture.} \quad We build our baseline convolutional neural network model with 3 convolutional layers (Conv2D), 2 MaxPooling layers, 4 Dropout layers, and 2 fully-connected layers. Table \ref{network} shows the network architecture used in our baseline network. This is a slightly modified version of the architecture used in \cite{CNN4}. We use L2 weight regularizer with a very small regularization hyperparameter $\alpha=0.0005$ in the first convolutional layer to reduce the overfitting. All Conv2D layers and the first fully connected dense layer have ReLU \cite{relu} as their activation function and the last dense layer has softmax as its linear activation function.

{\bf Training and evaluation.} \quad All networks were initialized using a default glorot uniform kernel initializer \cite{glorot} and trained using the Adam optimizer with its default parameters \cite{adam}. We train the networks on the modified Fashion-MNIST training set for a specific number of epochs which is determined using the early stopping method to prevent overfitting. We chose a batch size of 32 and trained the network for 30 times and reported the average of the test accuracy and its standard deviation for each data point that is shown in Fig.3 of the paper. As was mentioned in Sec.3 of the paper, we create three types of data sets based on Fashion-MNIST grey scale images. For the type A data set, we add to the images a black bezel of 6 pixel width by zero-padding. For the type B data set, we add noise only to the image and leave the bezel clear, as, e.g., in Fig.2a in the paper, and for the type C data set, we add noise to both the image and the bezel, as, e.g., in Figs.2b-f in the paper.

Let us look at the process of creating data set type B as an example. For the case where we want the images to consist of 50\% noise, first, we rescale each MNIST image with original pixel values ranging between 0 and 255 and its add-on black bezel  (pixel value zero) to an interval of [0.25, 0.75]. We rescale the images' pixel values so that when we add 50\% of the noise which is originally chosen to have the amplitude interval [-0.5,0.5], the pixel values of the noisy image do not overflow 1 or underflow 0. Similarly, to mention another example, if we want to have 70\% noise, we need to rescale original images and their bezel to an interval of [0.35,0.65]. Similarly, for other percentages of the noise, we do a corresponding rescaling, accordingly.

\begin{table}[h]\label{network}
  \centering
  \begin{tabular}{llll}
    \toprule
    \cmidrule(r){1-4}
    Layer     & Output shape     & Function &Number of parameters \\
    \midrule
    $\text{Conv2D}_1$ &  (None, 26, 26, 32)   & Convolution $3\times3$ & 320     \\
    $\text{MaxPooling2D}_1$     & (None, 13, 13, 32)   & Maxpool $2\times2$ &0      \\
    $\text{Dropout}_1$   & (None, 13, 13, 32) & Dropout 0.25 & 0  \\
    $\text{Conv2D}_2$ &  (None, 11, 11, 64)  & Convolution $3\times3$ & 18496    \\
    $\text{MaxPooling2D}_2$     & (None, 5, 5, 64)   & Maxpool $2\times2$ &0      \\
    $\text{Dropout}_2$   & (None, 5, 5, 64)  & Dropout 0.25 & 0  \\
    $\text{Conv2D}_3$ &  (None, 3, 3, 128)    & Convolution $3\times3$ & 73856     \\
    $\text{Dropout}_3$   & (None, 3, 3, 128)    & Dropout 0.4 & 0  \\
    $\text{Flatten}_1$ &   (None, 1152)   & Reshape to a vector  & 0    \\
    $\text{Dense}_1$     &(None, 128)  & Fully connected layer & 147584     \\
    $\text{Dropout}_4$   & (None, 128) & Dropout 0.3 & 0  \\
    $\text{Dense}_2$    & (None, 10)   & Fully connected layer & 1290     \\
    \bottomrule
  \end{tabular}
	\caption{Network architecture used in our experiments}
  \label{sample-table}
\end{table}

\section{Quantum field as a source of noise}

As we mentioned in Sections 3.2 and 4 of the paper, the low complexity noise that we used for our experiments is generated using sinusoidal noise panels with prefactors drawn from Gaussian probability distributions. We now show that this method of generating noise can be viewed as exactly simulating the quantum fluctuations of the amplitudes of a bandlimited massless neutral scalar field in the vacuum state.

Let $\phih(\bm x,t)$ denote a free, neutral, scalar quantum field in a flat spacetime of spatial dimension $d$, confined to a $d$-dimensional spatial hypercube of side length $L=1$ with Dirichlet boundary conditions $\phih(\bm x,t)|_{\text{boundary}}=0$. For generating 2-dimensional images, we  choose the dimension to be $d=2$. $\phih(\bm x,t)$ satisfies the Klein-Gordon equation
\begin{equation}
\label{eq:EOM}
    (\Box+m^2)\phih(\bm x,t)=0,
\end{equation}
the canonical commutation relations
\begin{align}
    \left[ \phih(\bm x,t),\phih(\bm x',t)\right]
    =
    \left[ \pih(\bm x,t),\pih(\bm x',t)\right]
    =0,\quad
    \left[ \phih(\bm x,t),\pih(\bm x',t)\right]
    =
    \ii \delta(\bm x-\bm x'),\label{eq:field_commutators}
\end{align}
and the self-adjointness condition 
 $   \label{eq:hermiticity}
    \phih^\dagger(\bm x,t)=\phih(\bm x,t)$,
where $\pih(\bm x,t):=\partial_t\phih(\bm x,t)$ and $m$ is the field's mass. For our simulations, we chose $m=0$. We express the field in terms of the modes of its Fourier sine expansion 
\begin{equation}\label{eq:Fourier_expansion}
    \phih(\bm x,t)=\sum_{\bm n\in\mathbb N^d}
    \phin(t)\bn(\bm x),
\end{equation}
where the $\bn(\bm x)$, with $\bm n = (n_1,\dots,n_d)\in\mathbb N^d$, are given by
\begin{equation}
\bn(\bm x)=2^{d/2}\prod_{i=1}^d 
\sin\left(\frac{n_i\pi x}{L}\right).
\end{equation}
The mode operators $\phin(t)$ then read
\begin{equation}\label{eq:phin}
    \phin(t)=\frac{1}{\sqrt{2\on}}
    \left(e^{\ii\on t}\ad n+e^{-\ii\on t}\a n\right).
\end{equation}
and we have, similarly, for their canonically conjugate operators, $\pin(t)$:
\begin{equation}\label{eq:pin}
    \pin(t)=\ii\sqrt{\frac{\on}{2}}
    \left(e^{\ii\on t}\ad n-e^{-\ii\on t}\a n\right).
\end{equation}
The frequency reads $\on=\sqrt{|\bm k_{\bm n}|^2+m^2}$, and the wave vectors $\kn$ are defined as
$    \kn:=\frac{\pi}{L}(n_1,n_2,\dots,n_d).$
The annihilation and creation operators $\a n$ and $\ad n$ obey
\begin{align}
    [\a n,\a m]=[\ad n,\ad m]=0,\quad
    [\a n,\ad m]=\delta_{\bm n,\bm m},\label{eq:CCR_a_ad}
\end{align}
and the vacuum state $\ket 0$ obeys 
 $   \a n\ket0 = 0 ~~ \text{for all } n.$ 
In addition to the infrared regularization provided by the box, we also impose an ultraviolet regularization that bandlimits the field by truncating the sum in Eq.\eqref{eq:Fourier_expansion}. In our simulations, we have $d=2$ and we truncated the sum at $n_i=5,15,22$, leading to $5^2,15^2$ and $22^2$ dimensional noise spaces, respectively. 

Let us now derive the probability distribution for field amplitude measurements in the vacuum state $\ket 0$. To this end, we utilize that, in Eq.\ref{eq:Fourier_expansion}, 
the field is decomposed into independent harmonic oscillators. The pairs of conjugate degrees of freedom $\phin(t)$ and $\pin(t)$ are self-adjoint and satisfy $[\phin(t),\hat\pi_{\bm n'}(t)]=\ii\delta_{\bm n,\bm n'}$. We choose a fixed time, such as $t=0$. 

In the $\phin$ eigenbasis, we have $\phin=\phinn$ and $\pin=-\ii\frac{\partial}{\partial\phinn}$, so that the condition $\a n\ket{0}=0$ becomes
\begin{equation}\label{eq:vacuum_ODE}
    \sqrt{\frac{\on}{2}}
    \left[
    \phinn+\frac{\ii}{\on}\left(
    -\ii\frac{\partial}{\partial\phinn}\right)
    \right]\psi_0(\phinn)=0,
\end{equation}
where we have one equation for each value of $\bm n$. These equations are solved to yield the normalized wavefunctions of the ground states of harmonic oscillators,
\begin{equation}\label{eq:vacuum_wavefcn}
    \psi_0(\phinn)=
    \left(\frac{\on}{\pi}\right)^{1/4}
    \exp\left(-\frac{\on}{2}\phinn^2\right).
\end{equation}
Thus when the quantum field is in its ground state, $\ket 0$, measurement outcomes of the amplitudes of each of the field modes $\phin$ are each independently following a Gaussian probability distribution function $|\psi_0(\phinn)|^2$ with mean zero and variance $1/2\on$. By drawing from this probability distribution and using Eq.\eqref{eq:Fourier_expansion}, we obtain a sample measurement of the field. This matches exactly the calculations of the low-complexity noise panels in our experiments.

\end{document}